\documentclass[aps,pra,amsmath,amssymb, reprint, superscriptaddress]{revtex4-1}
\usepackage{graphicx}
\usepackage{bm}
\usepackage{array}
\usepackage{dcolumn}
\usepackage{hyperref}
\usepackage{braket}
\usepackage{color}

\newcommand{\w}{3.25in}
\newcommand{\mz}{$m=0$}
\newcommand{\mpl}{$m=+1$}
\newcommand{\mmi}{$m=-1$}
\newcommand{\mplmi}{$m=\pm1$}
\newcommand{\ctwon}{$c_2\bar{n}$}

\newcommand{\overbar}[1]{\mkern 1.5mu\overline{\mkern-1.5mu#1\mkern-1.5mu}\mkern 1.5mu}

\newcommand{\comment}[1]{}

\newcommand{\Interferometer}[1][\w]{
\begin{figure*}[ht!]
\includegraphics*[width=5in,angle=0]{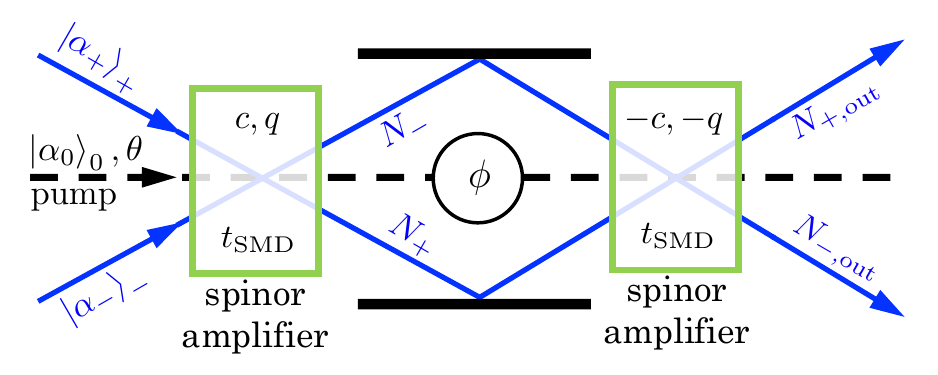}
\caption{\label{fig:Interferometer}(Color online) Configuration of the seeded spinor SU(1,1) interferometer. The initial spinor amplifier (first green box) mixes the pump coherent state $\ket{\alpha}_0$ (dashed black line) with the probe and conjugate coherent seeds $\ket{\alpha_\pm}_\pm$ (blue solid lines) depending on the relative spinor phase $\theta$. For a given set of input states, the amplifier gain is determined by the parameters for spinor dynamics $c,q$, and $t_\text{SMD}$ (defined in section \ref{sec:spinor}). A shift in the spinor phase within the interferometer $\phi$ is sensed when a second parametric amplifier (second green box) reverses the dynamics by negating the values of $c$ and $q$, and evolving the state for the same time $t_\text{SMD}$. The output of the interferometer is the number of atoms $N_{+,\text{out}}+N_{-,\text{out}}$.}
\end{figure*}
}

\newcommand{\AmpVsPhase}[1][\w]{
\begin{figure*}[ht!]
\includegraphics*[width=3.25in,angle=0]{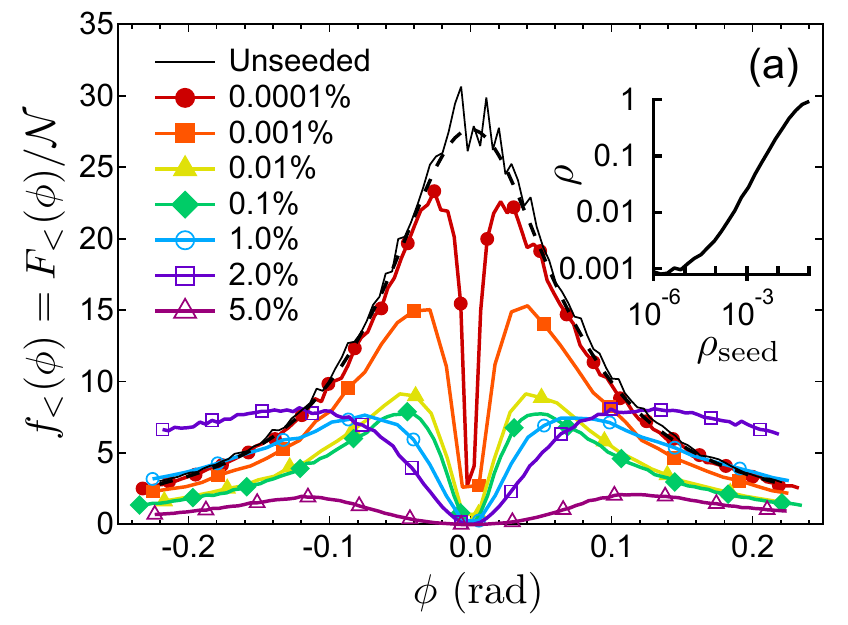}%
\includegraphics*[width=3.25in,angle=0]{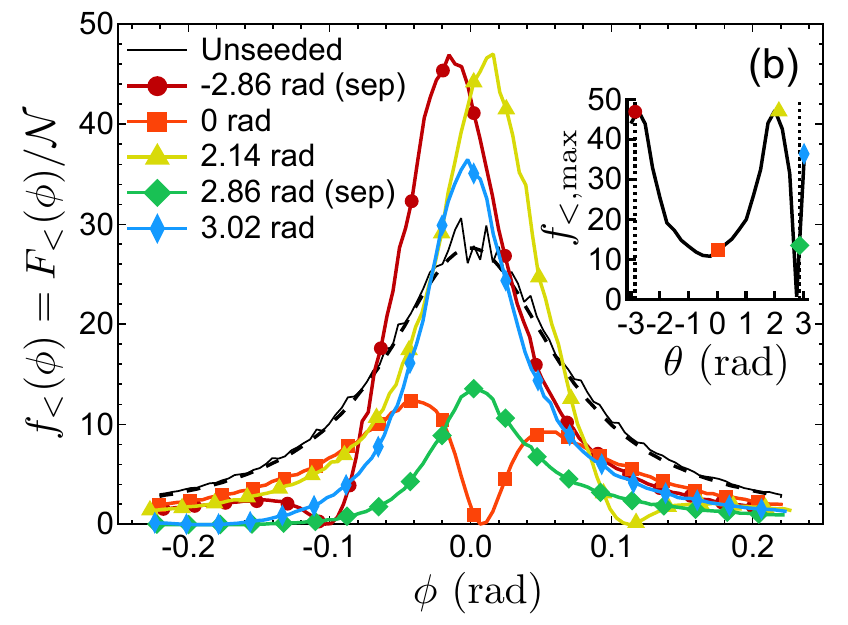}
\caption{\label{fig:AmpVsPhase}Lower bound on the scaled Fisher information obtained from TWA simulations versus interferometer phase $\phi$ for $t_\text{SMD} = 20\text{ ms}$ with (a) increasing percentages of single-sided seeding into \mmi, and with (b) increasing spinor phase for 0.1\% double-sided seeding. The dashed line shows $f(\phi)$ calculated from Eq. (\ref{eqn:FIBogo}) for the unseeded case. The inset to (a) is the amplified fraction after spin-mixing dynamics versus the initial seed fraction $\rho_\text{seed}$. The inset to (b) is $f_{<,\text{max}}$ versus initial spinor phase, with colored dots corresponding to the traces shown in the main figure. Vertical dotted lines indicate the phase of the separatrix (sep) $\theta_\text{sep}=\pm2.86\text{ rad}$ that divides oscillating phase and running phase spinor dynamics \cite{Zhang:2005fk}.}
\end{figure*}
}

\newcommand{\FIVsRho}[1][\w]{
\begin{figure*}[ht!]
\includegraphics*[width=3.25in,angle=0]{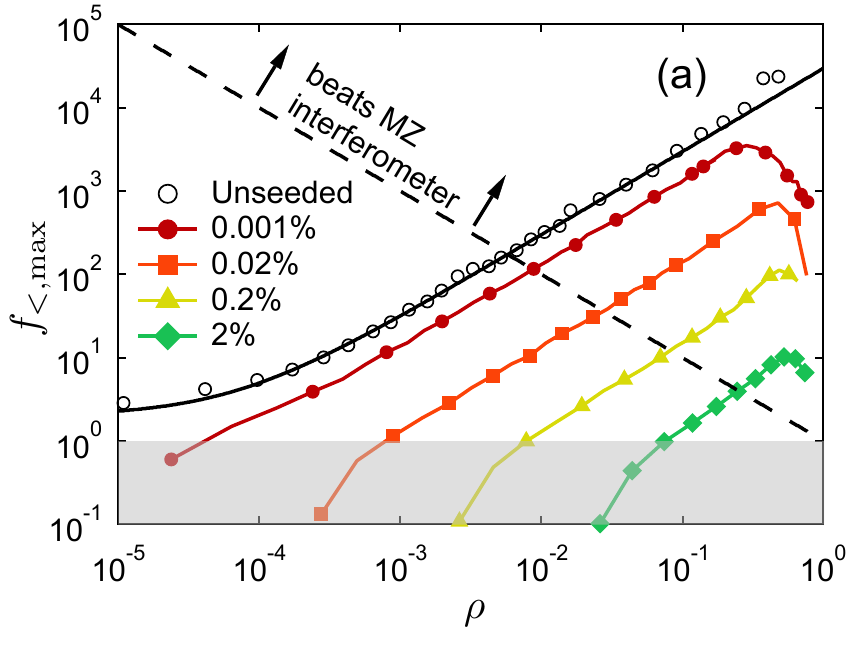}%
\includegraphics*[width=3.25in,angle=0]{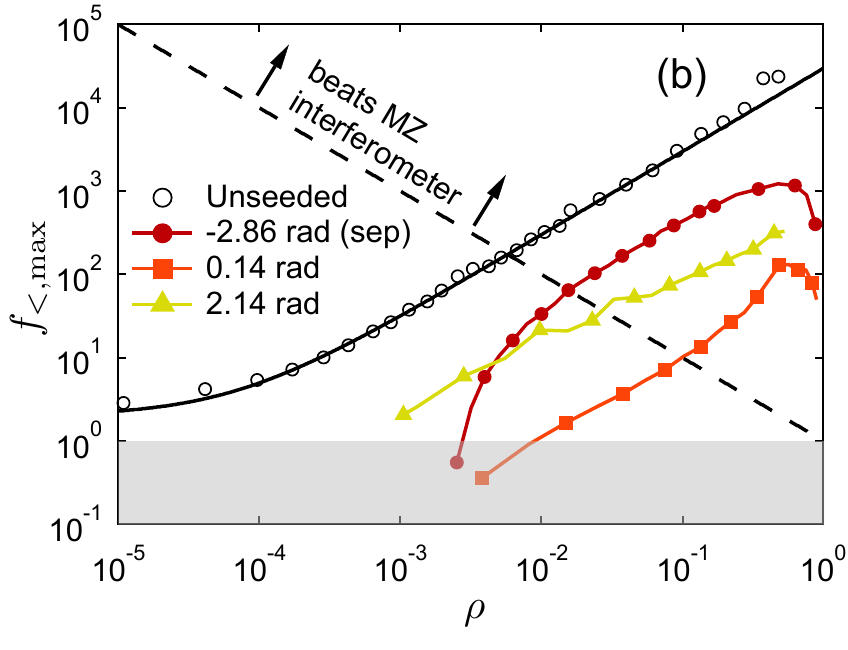}
\caption{\label{fig:FIVsRho}The lower bound on the scaled Fisher information versus $\rho$ with (a) increasing percentages of single-sided seeding into \mmi, and with (b) double-sided $0.1$\% seeding at various initial phases. The solid black line is $f_\text{Bog,max}=(\mathcal{N}+2)$, which is the prediction for the unseeded case (open circles). The grey shading indicates the region of scaled Fisher information that does not reach standard quantum limit $f_\text{SQL}=1$ (based on $\mathcal{N}$ atoms in the interferometer arms). The dashed line corresponds to $F= N$ (also $f=1/\rho$), which is the limit achievable using coherent states in a Mach-Zehnder interferometer having a total of $N$ particles. The phases of the separatrix are $\theta_\text{sep}=\pm2.86\text{ rad}$.}
\end{figure*}
}

\newcommand{\PhaseScan}[1][\w]{
\begin{figure}[ht!]
\includegraphics*[width=3.25in,angle=0]{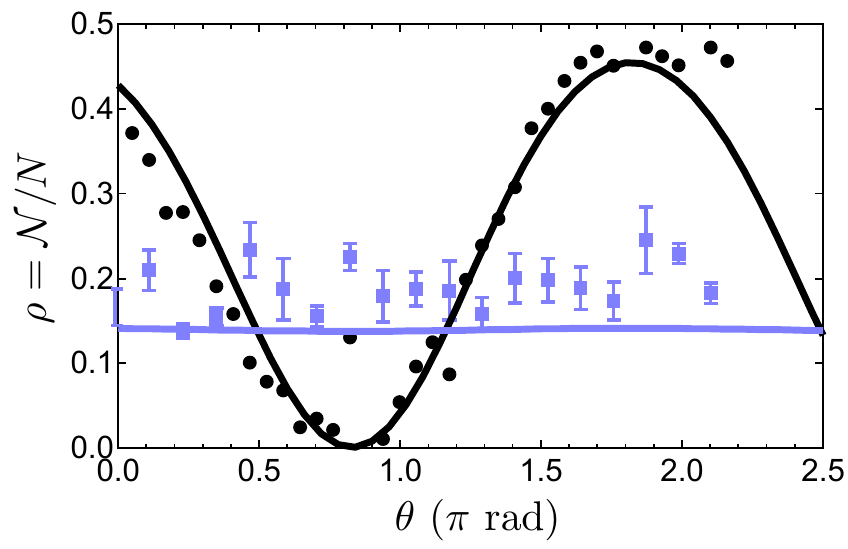}
\caption{\label{fig:phasescan}The atom fraction in \mplmi\ after 14 ms of evolution time versus the phase shift achieved by a microwave pulse. The amplifier is seeded with 1.0\% of the atoms in \mpl\ (blue squares) or with 1\% of the atoms in each of \mplmi\ (black circles) out of $N=3.6(5)\times10^4$ atoms. Solid lines are predictions from the single-mode spinor theory. Single-sided seeding represents multiple experiments and we quote the mean and standard deviation of the mean. Double-sided seeding represents one experiment per phase.}
\end{figure}
}

\newcommand{\AtomAmplifier}[1][\w]{
\begin{figure}[ht!]
\includegraphics*[width=3.25in,angle=0]{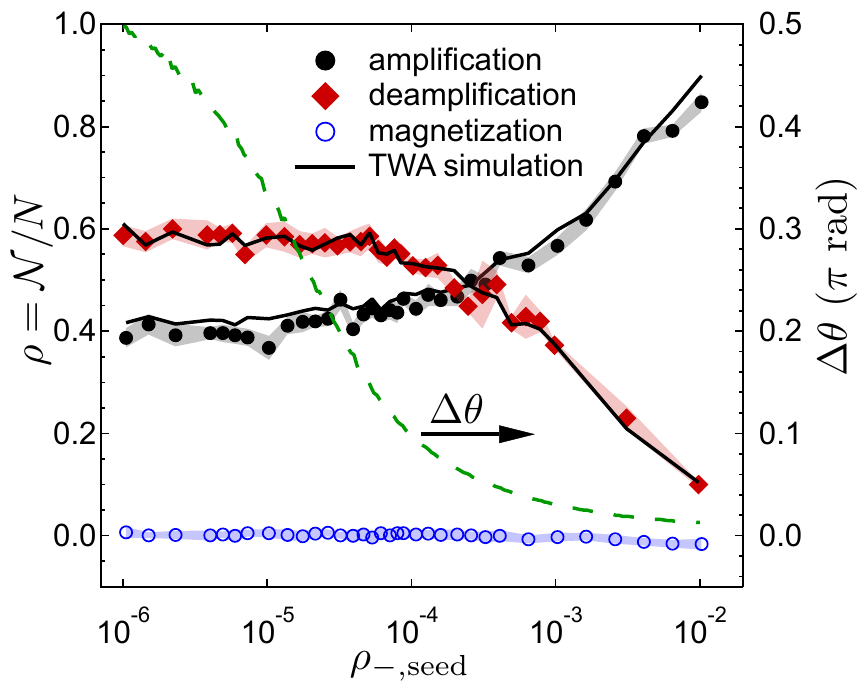}
\caption{\label{fig:AtomAmplifier}Experimental atom fraction after amplification for the seeded nonlinear amplifier versus seeded atom fraction in $m=-1$, with \mpl\ initially populated with a fraction $\rho_{+,\text{seed}}=0.01$ of the atoms. Solid black circles ($N=3.4(3)\times 10^4$, $q/h=-1$~Hz, $\theta=0.016$~rad, $t_\text{SMD}=27\text{ ms}$) have an initial spinor phase $\theta$ causing amplification, while red diamonds ($N=7.8(9) \times 10^4$, $q/h=-1.4$~Hz, $\theta=2.45$~rad, $t_\text{SMD}=25\text{ ms}$) have an initial phase causing deamplification. The shading indicates the standard deviation of the mean, and the solid black lines are results based truncated-Wigner approximation simulations. The green dashed line shows the decrease in the standard deviation ($\Delta\theta$) of the distribution of initial phases from the TWA simulations.}
\end{figure}
}

\begin{document}

\title{A spinor Bose-Einstein condensate phase-sensitive amplifier for SU(1,1) interferometry}

\author{J. P. Wrubel}
\affiliation{Department of Physics, Creighton University, 2500 California Plaza, Omaha, Nebraska 68178, USA}
\affiliation{Quantum Measurement Division, National Institute of Standards and Technology, and Joint Quantum Institute, NIST and University of Maryland, 100 Bureau Drive, Gaithersburg, Maryland 20899-8424, USA}
\author{A. Schwettmann}
\affiliation{Homer L. Dodge Department of Physics and Astronomy, The University of Oklahoma, 440 W. Brooks Street, Norman, Oklahoma 73019, USA}
\affiliation{Quantum Measurement Division, National Institute of Standards and Technology, and Joint Quantum Institute, NIST and University of Maryland, 100 Bureau Drive, Gaithersburg, Maryland 20899-8424, USA}
\author{D. P. Fahey}
\author{Z. Glassman}
\affiliation{Quantum Measurement Division, National Institute of Standards and Technology, and Joint Quantum Institute, NIST and University of Maryland, 100 Bureau Drive, Gaithersburg, Maryland 20899-8424, USA}
\author{H. K. Pechkis}
\affiliation{Department of Physics, California State University, Chico, CA, 95973, USA}
\affiliation{Quantum Measurement Division, National Institute of Standards and Technology, and Joint Quantum Institute, NIST and University of Maryland, 100 Bureau Drive, Gaithersburg, Maryland 20899-8424, USA}
\author{P. F. Griffin}
\affiliation{Department of Physics, University of Strathclyde, Glasgow G4 0NG, UK}
\affiliation{Quantum Measurement Division, National Institute of Standards and Technology, and Joint Quantum Institute, NIST and University of Maryland, 100 Bureau Drive, Gaithersburg, Maryland 20899-8424, USA}
\author{R. Barnett}
\affiliation{Department of Mathematics, Imperial College London, London, United Kingdom, SW7 2AZ}
\affiliation{Department of Physics, Joint Quantum Institute and Condensed Matter Theory Center, University of Maryland, College Park, Maryland 20742, USA}
\author{E. Tiesinga}
\author{P. D. Lett}
\affiliation{Quantum Measurement Division, National Institute of Standards and Technology, and Joint Quantum Institute, NIST and University of Maryland, 100 Bureau Drive, Gaithersburg, Maryland 20899-8424, USA}

\date{\today}

\begin{abstract}
The SU(1,1) interferometer was originally conceived as a Mach-Zehnder interferometer with the beam-splitters replaced by parametric amplifiers. The parametric amplifiers produce states with correlations that result in enhanced phase sensitivity. $F=1$ spinor Bose-Einstein condensates (BECs) can serve as the parametric amplifiers for an atomic version of such an interferometer by collisionally producing entangled pairs of $\ket{F=1,m=\pm1}$ atoms. We simulate the effect of single and double-sided seeding of the inputs to the amplifier using the truncated-Wigner approximation. We find that single-sided seeding degrades the performance of the interferometer exactly at the phase the unseeded interferometer should operate the best. Double-sided seeding results in a phase-sensitive amplifier, where the maximal sensitivity is a function of the phase relationship between the input states of the amplifier. In both single and double-sided seeding we find there exists an optimal phase shift that achieves sensitivity beyond the standard quantum limit. Experimentally, we demonstrate a spinor phase-sensitive amplifier using a BEC of $^{23}$Na in an optical dipole trap. This configuration could be used as an input to such an interferometer. We are able to control the initial phase of the double-seeded amplifier, and demonstrate sensitivity to initial population fractions as small as 0.1\%.
\end{abstract}


\maketitle
\section{Introduction}

Quantum coherent states of photons or matter are the workhorses for interferometric measurements because they can be made from large numbers of particles with well-defined phases. Although classical plane waves have no phase uncertainty, quantum coherent states of $N$ particles have an inherent phase uncertainty that limits the interferometer's sensitivity to the standard quantum limit (SQL) $\Delta\phi_\text{SQL}\propto 1/\sqrt{N}$. Caves \textit{et al.} \cite{Caves:1981ij,Caves:1982hd} realized that an interferometer exploiting quantum correlations between the input modes could surpass the SQL, reaching phase-sensitivities scaling with the Heisenberg limit $\Delta\phi\propto1/\mathcal{N}$, where $\mathcal{N}$ is the number of correlated particles. The challenge in achieving such high sensitivity is producing and maintaining highly non-classical correlated states of photons or atoms.

One solution to the problem of producing quantum correlations for interferometry is to replace the beamsplitters in a Mach-Zehnder interferometer with parametric amplifiers (Fig. \ref{fig:Interferometer}). Parametric amplifiers have a bright ``pump'' mode containing the vast majority of the particles, and amplify the seeded or unseeded ``probe'' and ``conjugate'' states. Such a parametric amplifier may be constructed from a single-mode Bose-Einstein condensate within the Zeeman-split ground-state hyperfine manifold $\ket{F,m}$, which is called a spinor parametric amplifier \cite{Peise:2015gr,Bucker:2012dm,Scherer:2010hs}.

We consider the hyperfine states $\ket{F=1,m}$ having complex amplitudes represented by the spinor $\xi=(\xi_{+1,},\xi_{0},\xi_{-1})^T$, with $\xi_m=\sqrt{\rho_m}e^{i\theta_m}$, mean fractional populations $\rho_m=N_m/N$ and phases $\theta_m$. The initial state of the BEC of $N$ atoms is represented by a product of the coherent states $\ket{\alpha_m}_m=\ket{\sqrt{N}\xi_m}_m$. In the spinor parametric amplifier $\ket{\alpha_0}_0$ serves as the pump and $\ket{\alpha_\pm}_{\pm}$ provide the probe and conjugate modes.

\Interferometer

Parametric amplifiers can be modeled analytically using a linear ``undepleted pump'' approximation, resulting in an interferometer design with SU(1,1) symmetry \cite{YURKE:1986uf}. When the input parametric amplifier produces an average of $\mathcal{N}=(N_++N_-)\gg1$ particles in the probe and conjugate states, then the minimum phase-sensitivity of the interferometer $\Delta\phi=1/\sqrt{\mathcal{N}(\mathcal{N}-2)}$ approaches the Heisenberg limit \cite{YURKE:1986uf}. This scaling with the number of atoms in the probe and conjugate states $\mathcal{N}$ is not identical to scaling with the total number of particles $N=N_\text{pump}+\mathcal{N}$. This SU(1,1) interferometer has been explored in various systems theoretically \cite{YURKE:1986uf, Plick:2010fw, Ou:2012eq,Marino:2012vh, Gabbrielli:2015kg,Szigeti:2017br}, and the unseeded interferometer has been explored experimentally \cite{Gross:2010jn,Ma:2015hn,Cheung:2016tq,Linnemann:2016iu,Anderson:2017bc,Manceau:2017cm,Hudelist:2014di}.

The atomic spinor SU(1,1) interferometer is sketched in Fig. \ref{fig:Interferometer}, where initially $0-1$\% of the atoms are in the probe and conjugate states \mplmi, with the remainder in the pump state \mz. After the input amplifier, a phase shift $\phi$ is applied to the relative spinor phase $\theta=\theta_{+1}+\theta_{-1}-2\theta_0$, and the amplification is reversed in a second parametric amplifier. The measured output of the interferometer is the sum of the atoms in the $m=\pm1$ levels $\mathcal{N}_\text{out}=N_{+,\text{out}}+N_{-,\text{out}}$.

Optical parametric amplifiers typically operate well within the undepleted pump approximation due to the weakness of four-wave mixing and down conversion processes. Study of the seeding of the initial probe and conjugate states of the optical SU(1,1) interferometer showed that sensitivity beyond the standard quantum limit is lost at $\phi=0$, which would otherwise be the phase yielding maximum sensitivity in the unseeded case. By operating the interferometer away from $\phi=0$, the interferometer could still surpass the SQL \cite{Plick:2010fw,Marino:2012vh,Anderson:2017bb}.

In contrast to optical amplifiers, spinor parametric amplifiers in a single spatial mode rapidly deplete the finite pump resource, transferring a majority of the atoms to the probe and conjugate states, and breaking the linear approximation. Simulations of this nonlinear spinor amplifier as an input to the SU(1,1) interferometer showed that Heisenberg-limited sensitivity was retained, although the effects of initial seeding of the probe and conjugate states was not considered \cite{Gabbrielli:2015kg}. Extending these results, we simulate the nonlinear spinor amplifier as an input to the SU(1,1) interferometer, including the cases of small initial coherent seeds in one or both of the probe and conjugate states, which has not previously been considered.

Demonstrating a spinor SU(1,1) interferometer capable of surpassing the standard quantum limit is a challenging task. The best results with a vacuum-seeded spinor SU(1,1) interferometer demonstrated sensitivity somewhat better than the SQL: $\Delta\phi\approx\Delta\phi_\text{SQL}/(1.6\pm0.5)$ \cite{Linnemann:2016iu}. Even in this vacuum-seeded case, it was observed that the sensitivity of the interferometer was suppressed near $\phi=0$, where the ideal interferometer (without losses) would have been maximally sensitive.

The difficulty presented by the full SU(1,1) interferometer of Ref. \cite{YURKE:1986uf} is that at the operating point with maximal theoretical sensitivity, the mean and variance of the detected particles at the output, $\mathcal{N}_\text{out}$, go to zero leaving the measurement susceptible to noise from losses \cite{Marino:2012vh}, or imperfect reversibility \cite{Linnemann:2016iu}. We propose that a third reason for loss in sensitivity of the spinor SU(1,1) interferometer is imperfect state preparation. If even very small fractions of atoms are present in \mplmi, then the initial probe and conjugate states are not exact vacuum-states and the effects of these imperfections must be considered.

Here, we simulate the nonlinear spinor SU(1,1) interferometer using the truncated-Wigner approximation \cite{Blakie:2008is,Polkovnikov:2010dm} in a single spatial mode for the case of coherent single and double-sided seeding of the input states. For single sided seeding, we find a decrease in sensitivity at $\phi=0$ analogous to the optical interferometer \cite{Plick:2010fw,Marino:2012vh}, while surpassing the standard quantum limit at other phases.

We also simulate the case of double-sided seeding of the input amplifier, resulting in finite populations in both output modes, which are therefore less susceptible to measurement noise. Our simulations show that a nonlinear SU(1,1) interferometer initiated using double-sided seeding of the \mplmi\ states can significantly surpass the standard quantum limit. For particular values of the initial spinor phase $\theta$ of the phase-sensitive amplifier with fixed evolution time, we find a maximum sensitivity not only surpassing the SQL, but also surpassing the expectation for vacuum-seeding.

Experimentally, we demonstrate the phase sensitivity of a spinor parametric amplifier that could be used to construct such an SU(1,1) interferometer using a $^{23}$Na BEC in a crossed optical dipole trap \cite{Pechkis:2013vg}. The BEC of $N\approx3.5\times 10^4$ atoms meets the single spatial-mode criterion so that the spatial wave-function can be ignored in the dynamics. We show a three sigma change in the amplifier output when one input to the phase-sensitive amplifier is seeded with a fraction of just $10^{-3}$.

\section{entanglement in spinor BECs}\label{sec:spinor}
Entanglement in spinors has been produced both by transferring the atoms into an unstable $m=0$ level that spontaneously decays into entangled pairs of $m=\pm1$ atoms \cite{Gross:2011ef,Strobel:2014eg,Peise:2015gr}, as well as by adiabatic passage through quantum phase transitions \cite{Luo:2017jo}. The pairs of atoms produced from an unstable spinor state in the undepleted pump regime are analogous to the pairs of photons produced in downconversion or four-wave mixing \cite{Peise:2015gr}. The atomic state that results from this process in the linear approximation is a two-mode vacuum-squeezed state.

Spontaneous emission of correlated $m=\pm1$ atom pairs can be induced in a sodium BEC when the $\ket{1,0}$ level has a slightly higher energy than the average energy of the $\ket{1,\pm1}$ levels. The gain of the parametric amplifier can be controlled by varying the time $t_\text{SMD}$ allowed for this spin-mixing dynamics (SMD) to proceed, from very short times for which a linear Bogoliubov approximation is valid, to very long times where this approximation breaks down and the depletion of the pump state must be taken into account. During spin-mixing dynamics magnetization-conserving spin-flip collisions convert a pair of $m=0$ atoms into a correlated pair of $m=\pm1$ atoms and vice-versa.

The most significant difference between entangled photons and entangled atoms is the interaction energy $c=c_2\bar{n}$ between the atoms that introduces nonlinearity into the evolution, where $c_2=2\pi\hbar^2(a_2-a_0)/3\mu$, the $a_F$ are the scattering lengths for two atoms with total spin $F$, $\mu$ is the reduced mass of two atoms, and $\bar{n}$ is the mean density of the BEC \cite{Zhang:2005fk}. In our crossed optical dipole trap, the atoms remain in a single spatial wavefunction for the duration of the experiment, so that every pair of atoms created further increases the rate of pair production, exponentially increasing the number of atom pairs for short times.

When the spin-components of the BEC all share a single spatial wavefunction we can make the single-mode approximation (SMA). The SMA leads to a simplification of the dynamics of the mean-field spinor $\xi=(\xi_{+1},\xi_0,\xi_{-1})^T$, where $\ket{\xi_m}=\ket{\sqrt{\rho_m}e^{i\theta_m}}$, down to two dynamical variables: $\rho_0$, which is the fractional population in \mz, and the spinor phase $\theta=\theta_{+1}+\theta_{-1}-2\theta_0$ \cite{Zhang:2005fk}. The dynamics is a function of the spin-dependent interaction energy $c$, and the quadratic Zeeman energy $q=(E_-+E_+-2E_0)/2$, where $E_m$ is the energy of the each hyperfine state $\ket{1,m}$. The magnetization $M\equiv\rho_{+}-\rho_{-}$ is conserved in the SMA.

Spinor parametric amplifiers typically avoid the nonlinear regime by restricting the dynamics to short times so that only a very small fraction of atoms produce correlated pairs \cite{Kruse:2016bm,Linnemann:2016iu}. In the linear or ``undepleted pump" regime, where the initial population in $m=0$ has not changed significantly, the quantum spin-mixing dynamics is modeled by making a linear Bogoliubov approximation. The Bogoliubov approximation remains valid when the fraction of entangled atoms produced from spin-mixing dynamics (SMD) is $\rho_{\mathcal{N}}=\rho_++\rho_-\lesssim 0.01$. Although an SU(1,1) interferometer may have phase sensitivity that scales with the Heisenberg limit $\Delta\phi\propto1/\mathcal{N}$, $\mathcal{N}$ is small and the absolute sensitivity is limited. Nevertheless, an enhancement of 2.05 dB over the SQL was achieved in a measurement of the hyperfine clock transition ($\ket{F,m}=\ket{1,0}\rightarrow\ket{2,0}$) in a spinor BEC using an average of just 0.75 entangled atom pairs \cite{Kruse:2016bm}.

Here we consider the case of producing large fractions of entangled $m=\pm1$ atoms such that the pump is significantly depleted. This nonlinear SU(1,1) interferometer is in principle capable of significantly enhanced absolute sensitivity since almost all the pump atoms participate in the measurement \cite{Gabbrielli:2015kg}. The challenge introduced by the nonlinear SU(1,1) interferometer is negating the sign of the interaction energy $c$ and quadratic Zeeman shift $q$ to achieve perfect reversal of the spin dynamics. Although an a.c. Stark shift may be used to reverse the sign of $q$, it is extremely difficult to reverse the sign of the spinor interaction energy $c$. Feshbach resonances \cite{Chin:2010vi} can be used to modify atomic interactions, but have limited applicability to the spinor system because of the need for large magnetic fields or short experiment times. Within the Bogoliubov approximation the reversal may be achieved without negation of $c,q$ using an applied phase shift $\phi$ inside the interferometer \cite{Linnemann:2016iu}.

\section{Theory and Simulations of the Seeded Nonlinear SU(1,1) Spinor Interferometer}

An SU(1,1) interferometer is structurally similar to the Mach-Zehnder interferometer. In an unseeded SU(1,1) spinor interferometer, the input is a coherent state with all atoms in $\ket{\alpha_0}_0$. Spin-mixing dynamics (SMD) of the atoms in the first amplifier produces pairs of entangled $m=\pm 1$ atoms in a time $t_{\rm SMD}$. Unlike in a photon amplifier, the atoms are always within the ``gain'' medium, and nearly all atoms can be transferred out of $m=0$. The number of atoms in $m=\pm 1$ within the interferometer after the spin-mixing dynamics is ${\cal N}$. At this point a phase shift $\phi$ is applied and the spin-mixing dynamics is reversed using a second spinor amplifier. Following Ref.~\cite{Gabbrielli:2015kg} we treat the optimal case of exactly reversing the Hamiltonian by mapping $c \to -c$ and $q\to -q$. The spinor is allowed to evolve back for the same amount of time $t_{\rm SMD}$, and the output of the interferometer is the summed atom number in $m=\pm1$, $\mathcal{N}_{\rm out} = N_{+,\rm out} + N_{-,\rm out}$.

The quantum Hamiltonian of a spin-one Bose-Einstein condensate occupying a single spatial mode \cite{Zhang:2005fk,Ho:1998vi,Law:1998ty} is $H = c/(2N) \vec F\cdot \vec F - qa_0^\dagger a_0$ , where $\vec F = \sum_{mm'} a^\dagger_m {\vec F}_{mm'}a_{m'}$ is the total spin operator, ${\vec F}_{mm'} = (F^x_{mm'},F^y_{mm'},F^z_{mm'})$, and $F^{x,y,z}_{mm'}$ are matrix elements of the $x$, $y$, and $z$ spin-one matrices. Finally, $a^\dagger_m$ ($a_m$) are the creation (annihilation) operators for an atom in spin state $m= +1, 0, -1$ and in the relevant single spatial mode. The mean field language implicitly used in the first two sections follows from a unitary transformation of $ H$ under the displacement operator $D= \prod_{m=-1}^1 \exp(\alpha_m a^\dagger_m-\alpha_m^*a_m)$. The part of the transformed Hamiltonian that is independent of the creation and annihilation operators is minimized with respect to the $\alpha_m$, and only terms that are up to quadratic in the creation and annihilation operators are kept. The operator-independent terms of the transformed Hamiltonian can be recognized as the classical or mean-field spinor Hamiltonian for the $\alpha_m$ and $\alpha^*_m$ or equivalently $\rho_m$ and $\theta_m$. The quadratic terms correspond to the Bogoliubov Hamiltonian. Its role is discussed in the next section.

For spin-1 alkali-metal atoms in the presence of a small homogeneous magnetic field the parameter $q$ corresponds to the quadratic Zeeman shift and is non-negative. For sodium atoms $c>0$ and the ground state of the mean field Hamiltonian at zero magnetization ($M=0$) is a coherent state with all atoms in $m=0$, i.e. $\rho_0=1$. The application of a microwave field can change the sign of $q$ and for $-2c < q < 0$ \cite{Gerbier:2006jt} this causes a dynamical instability as the classical Hamiltonian has a saddle point at $\rho_0=1$. Because of this instability pairs of $m = 0$ atoms will spontaneously scatter into pairs of correlated atoms in the $m=\pm1$ states. The amplification process is called phase-insensitive because the spinor phases $\theta_{+1}$, $\theta_{-1}$ and $\theta_0$ are ill- or un-defined. Even if only one of the $m=\pm1$ states is seeded so that one of $\rho_{\pm1}>0$ and the corresponding phase is defined, the amplification remains phase-insensitive, since the initial spinor phase $\theta=\theta_{+1}-\theta_{-1}-2\theta_0$ is still undefined.

In a double-seeded SU(1,1) interferometer, the input has most of the atoms in the $m=0$ state, but also small coherent populations in \mplmi. Hence, the initial phase $\theta$ of the spinor is well-defined. This initial phase controls the subsequent amplification or de-amplification of pairs of atoms and characterizes the phase-sensitive amplifier.

The figure of merit for an interferometer is its phase sensitivity $\Delta\phi$. We estimate $\Delta\phi$ from simulations of the mean $\mathcal{N}_\text{out}$ and variance $(\Delta\mathcal{N}_\text{out})^2$ of the total number of atoms in \mplmi\ at the output of the interferometer as a function of the phase shift $\phi$ within the interferometer. Using error propagation to estimate the phase sensitivity gives
\begin{equation}
\label{eqn:phi_ep}
\Delta\phi=\Delta\mathcal{N}_\text{out}/\left(\frac{d\mathcal{N}_\text{out}}{d\phi}\right).
\end{equation}

The Fisher information $F(\phi)$ can be used to quantify the presence of useful entanglement in a system. $F(\phi)$ is defined in terms of the conditional probability $P(\mathcal{N}_\text{out}|\phi)$ of obtaining $\mathcal{N}_\text{out}$ atoms at the output given an interferometer shift $\phi$ \cite{Gabbrielli:2015kg}
\begin{equation}
\label{eqn:FI}
F(\phi)=\sum_{\mathcal{N}_\text{out}=0}^\infty \frac{1}{P(\mathcal{N}_\text{out}|\phi)}\left(\frac{dP(\mathcal{N}_\text{out}|\phi)}{d\phi}\right)^2.
\end{equation}
The maximum of the Fisher information $F_\text{max}\equiv\max [F(\phi)]$ is achieved at an operating phase shift $\phi_\text{opt}$. If the measurement scheme chosen is not the optimal one, then $F_\text{max}$ underestimates the quantum Fisher information $F_Q$. Whereas $F_\text{max}$ is the best case for the SU(1,1) measurement scheme, $F_Q$ is maximized over the set of all possible quantum measurements \cite{Braunstein:1996ke,Pezze:2016wo}. The inequality $F_Q>\mathcal{N}$ provides a condition ``sufficient for entanglement and necessary and sufficient for entanglement useful for quantum metrology'' \cite{Pezze:2016wo}.

The Fisher information can be measured experimentally using the squared Hellinger distance \cite{Strobel:2014eg}; however, this technique does not work for simulations using the truncated-Wigner approximation as the distributions are broadened by quantum noise. We instead calculate $1/(\Delta\phi)^2$, which provides a lower-bound for the Fisher information $F(\phi)\ge F_<(\phi)=1/(\Delta\phi)^2$ \cite{Gabbrielli:2015kg}. The ultimate phase-sensitivity produced by $n$ repeated measurements is calculated by the Cram\'{e}r-Rao lower bound $\Delta\phi_\text{CR}=1/\sqrt{nF(\phi)} \le \Delta\phi/\sqrt{n}$.

The standard quantum limit for the Fisher information is $F_\text{SQL}=\mathcal{N}$, while the Heisenberg limit for large $\mathcal{N}$ is $F_\text{H}=\mathcal{N}^2$. We define the scaled Fisher information $f(\phi)\equiv F(\phi)/F_\text{SQL}=F(\phi)/\mathcal{N}$. With this definition the SQL corresponds to a scaled Fisher information $f_\text{SQL}=1$, and the Heisenberg limit is $f_\text{H}=\mathcal{N}$. Scaling the Fisher information in this way by the number of atoms in \mplmi\ after time $t_\text{SMD}$ gives an appropriate measure of entanglement in the system. If, however, we ask whether the interferometer performs better than a Mach-Zehnder interferometer with coherent state inputs, then we should scale the Fisher information by the total number of atoms in the BEC, $N$, instead of $\mathcal{N}$.

\subsection{Bogoliubov approximation for the phase-sensitive spinor amplifier}
In this subsection we expand on the linear Bogoliubov theory to include the cases of single and double-sided initial seeding. These situations correspond to the phase-insensitive amplifier (PIA), and phase-sensitive amplifier (PSA) respectively. We use the Bogoliubov approximation to derive an expression for the early-time evolution of the spinor that is analogous to the phase-sensitive amplification of photons in a nonlinear medium. The Bogoliubov approximation simplifies the Hamiltonian by considering only small deviations from the initial state.

In the photon case, a strong pump laser beam interacts with two weak beams, the probe and conjugate. All states are initially coherent states, and the total number of photons in the amplified probe and conjugate beams has four terms. One term is the initial populations, a second term is due to spontaneous emission, a third term is due to phase-insensitive gain, and a fourth term is due to phase-sensitive gain \cite{Marino:2012vh}.

For the single-mode spinor BEC, we derive a similar equation by using a Bogoliubov expansion of the \mplmi\ states, while assuming that the much larger number of atoms in \mz\ is a constant classical field. This has been done previously for the phase-insensitive case \cite{Gabbrielli:2015kg,Pedersen:2014ja}, but here we give the result for a phase-sensitive spinor amplifier, beginning with coherent states in $\ket{\alpha_\pm}_\pm$ with average initial number of atoms $\overbar{N}_\pm$, and initial phase $\overbar{\theta}$ (the bar indicates initial values).

The total number of atoms in \mplmi\ during amplification, $\mathcal{N}(t)$, in the Bogoliubov approximation is
\begin{widetext}
\begin{eqnarray}\label{eqn:bogoamp}
\mathcal{N}(t)=&&(\overbar{N}_++\overbar{N}_-)+2\left(\frac{c_2\bar{n}}{\hbar\omega}\right)^2\sinh^2(\omega t)+2(\overbar{N}_++\overbar{N}_-)\left(\frac{c_2\bar{n}}{\hbar\omega}\right)^2\sinh^2(\omega t)\nonumber\\
&&+4\sqrt{\overbar{N}_-\overbar{N}_+}\frac{c_2\bar{n}}{\hbar\omega}\sinh (\omega t)\times\left(\cosh(\omega t) \sin\overbar{\theta}+\frac{c_2\bar{n}+q}{\hbar\omega}\sinh(\omega t) \cos\overbar{\theta}\right).
\end{eqnarray}
\end{widetext}
The instability rate $\omega(q<0)\equiv\sqrt{|q|(2 c_2\bar{n}-|q|)}/\hbar$ is a maximum when the effective quadratic Zeeman shift is $q=-c_2\bar{n}$. As for the optical case, we identify four terms in the amplification process: initial population, spontaneous emission, phase-insensitive gain, and phase-sensitive gain. When $q=-c_2\bar{n}$ the term proportional to $\cos\overbar{\theta}$ in Eq. (\ref{eqn:bogoamp}) is zero, and the atom and photon amplifiers behave in closely analogous ways. The most distinct difference between the optical and spinor parametric amplifiers is the ability to nearly completely deplete the pump in the spinor amplifier.

\subsection{Fisher information}
The Fisher information of the unseeded SU(1,1) interferometer in the Bogoliubov approximation \cite{Gabbrielli:2015kg} is
\begin{equation}\label{eqn:FIBogo}
F_\text{Bog}(\phi)=\frac{\mathcal{N}(\mathcal{N}+2)}{\mathcal{N}(\mathcal{N}+2)\sin^2(\phi/2)+1}\cos^2(\phi/2),
\end{equation}
where $\mathcal{N}=N_+(t_\text{SMD})+N_-(t_\text{SMD})$ is the number of atoms in \mplmi\ after the initial amplification due to spin-mixing dynamics. Within this approximation, $F_\text{Bog,max}=F_\text{Bog}(\phi_\text{opt}=0)=\mathcal{N}(\mathcal{N}+2)$, or a scaled Fisher information of $f_\text{Bog,max}=(\mathcal{N}+2)$, with $\mathcal{N}>0$.

\subsection{Truncated-Wigner approximation}
We numerically estimate the Fisher information both within the Bogoliubov approximation (short $t_\text{SMD}$) and beyond to large depletion of \mz\ (long $t_\text{SMD}$) by calculating $\Delta\phi$ (Eq.~(\ref{eqn:phi_ep})) using the truncated-Wigner approximation to simulate the spinor wavefunction.

The truncated-Wigner approximation (TWA) improves upon the Gross-Pitaevskii Equation (GPE) method by allowing for the inclusion of quantum fluctuations \cite{Blakie:2008is,Polkovnikov:2010dm}. Briefly, the TWA simulations are averages over many GPE simulations with initial noise corresponding to the Wigner transform of the initial state. This noise has a form such that the short time dynamics agrees with that of Bogoliubov theory. We simulate $10^4$ trajectories to achieve good statistics on both the average values and standard deviations. Using the TWA we obtain the average number of atoms at the output of the interferometer $\mathcal{N}_\text{out}(\phi)$ and its uncertainty $(\Delta\mathcal{N})_\text{out}(\phi)$ as a function of interferometer phase. We calculate the derivative $d\mathcal{N}_\text{out}/d\phi$ for each particular instance of the random initial conditions using a symmetric phase step of $10^{-6}$ rad. In order to avoid the limits of the precision of floating point numbers for the small changes in atom number caused by the $10^{-6}$ rad change in phase, we take the derivative of each iteration (for instance, in Eq. (\ref{eqn:phi_ep})) before averaging instead of taking the derivative of the averaged quantities. With infinite precision these two methods would yield the same result. Note that when simulating using the TWA it is important to compensate the raw averages and variances for Weyl ordering of the operators \footnote{\label{WeylNote}The average number of atoms at the output of the interferometer $\mathcal{N}_\text{out}(\phi)$ and its variance $(\Delta \mathcal{N})^2_\text{out}(\phi)$ must be corrected from the raw TWA simulation values because of the offset due to Weyl ordering of the operators. To compensate for Weyl ordering, we would need to subtract half an atom per measured population, or 1 atom from $(\mathcal{N})_\text{TWA,final}$ in the limit of an infinite number of simulation averages. To account for fluctuations due to a finite number of averages we actually subtract $\mathcal{N}_\text{TWA,initial}-\overbar{\mathcal{N}}_\text{initial}$, where $\overbar{\mathcal{N}}_\text{initial}$ is the average number of atoms in the initial coherent seed in $\ket{\alpha_\pm}_\pm$. $\mathcal{N}_\text{TWA,initial}$ is larger than this value by 1 on average. Similarly, the simulated variance should be reduced by $(\Delta \mathcal{N})_\text{TWA,initial}^2-\overbar{\mathcal{N}}_\text{initial}$. Taken together, these shifts have the result of guaranteeing that the populations and variances of the initial coherent states both have the correct value, $\overbar{\mathcal{N}}_\text{initial}$.}.

\subsection{Simulation results}

The parameters used for the simulation are typical of a sodium spinor BEC with $N=3\times10^4$ atoms, $c/h=25\text{ Hz}$, and $q/h=-1\text{ Hz}$. The results do not depend significantly on $c$ or $q$ so long as comparisons are made with the same $\mathcal{N}$. The number of atoms is increased by increasing the time for spin-mixing dynamics $t_\text{SMD}$ up to about 100 ms or until $\rho_0$ is nearly depleted.

We estimate the Fisher information with $F_<(\phi) =1/(\Delta\phi)^2 \le F(\phi)$, and the scaled Fisher information with $f_<(\phi)=F_<(\phi)/\mathcal{N}$. Figure \ref{fig:AmpVsPhase}(a) shows a plot of $f_<(\phi)$ for various fractions of single-side seeding in \mmi\ with $t_\text{SMD}=20\text{ ms}$. For vacuum seeding the lower bound from the TWA simulation agrees well with the Bogoliubov theory, Eq. (\ref{eqn:FIBogo}). The TWA simulation is most sensitive to noise due to random quantum fluctuations near $\phi=0$ where $\Delta\mathcal{N}_\text{out}$ and $d\mathcal{N}_\text{out}/d\phi$ both approach zero.

\AmpVsPhase

The effect of even very small seed fractions is quite pronounced. For a fraction as small as $10^{-6}$, $f_<(\phi)$ is significantly suppressed at $\phi=0$. As the fractional seed increases, the narrow dip gets wider, and the maximum of $f_<(\phi)$ decreases until saturating near 0.1\% seeding. For 2.0\% seeding the amplified fraction of atoms is $\rho=\mathcal{N}/N\approx 0.5$, and for larger seed fractions the peak decreases as we enter the heavily depleted regime.

An important consequence of the increasing seed fraction $\rho_\text{seed}=\mathcal{N}_\text{seed}/N$ is the increased gain of the amplifier for fixed $t_\text{SMD}$, resulting in significantly larger numbers of atoms $\mathcal{N}$ after the initial amplification. This is demonstrated in the inset to Fig. \ref{fig:AmpVsPhase}(a) that shows the amplified fraction $\rho$ versus $\rho_\text{seed}$ for $t_\text{SMD}=20$ ms. Increasing the seed fraction also increases the fraction of atoms produced by the amplifier, effectively increasing the gain of the amplifier.

Bogoliubov theory suggests that the width of $f(\phi)$ versus $\phi$ should decrease as $\mathcal{N}$ increases, but this is not what is observed. Instead the full-width at half-maximum (FWHM) of $f_<(\phi)$ in Fig. \ref{fig:AmpVsPhase}(a) is nearly constant and similar to the FWHM in the unseeded case, whereas the width of the dip at $\phi=0$ does increase with $\mathcal{N}$. The deep dip at $\phi=0$ demonstrates that even small initial coherent populations suppress the Fisher information, as the mean and variance no longer approach zero. Despite the loss of sensitivity at $\phi=0$, there exists instead an optimal value of $\phi$ such that $f_{<,\text{max}}=f_<(\phi_\text{opt})$.

\FIVsRho

Another way of viewing the effects of single-sided seeding is to plot the lower bound on the scaled Fisher information $f_{<,\text{max}}=\max [f_<(\phi)]$ versus $\rho=\mathcal{N}/N$ (see Fig.~\ref{fig:FIVsRho}(a)). For a fixed seed fraction we increase the spin-mixing time $t_\text{SMD}$ until $\rho\approx1$ where the pump is nearly depleted. It is observed in Fig.~\ref{fig:FIVsRho}(a) that increasing the seed fraction causes $f_{<,\text{max}}$ to decrease. Even with just $0.02$\% seed, $f_{<,\text{max}}$ has decreased by about a factor of 30 compared to the unseeded interferometer, suggesting that imperfect state preparation would have a significant impact on the sensitivity of the interferometer. As $\rho$ increases, the scaled Fisher information peaks at a value orders of magnitude above the standard quantum limit before decreasing when $\rho\rightarrow 1$. The dashed line in Fig. \ref{fig:FIVsRho}(a) indicates $F=N$, or $f=1/\rho$. For $f_{<,\text{max}}$ above this line, the interferometer is more sensitive than a Mach-Zehnder interferometer with coherent inputs using all $N$ atoms in the BEC. This is the limit where the nonlinear spinor SU(1,1) interferometer has truly improved measurement precision by exploiting quantum correlations.

Seeding the initial state has the advantage that the output of the interferometer has approximately the same number of atoms as in the initial seed, $\mathcal{N}_\text{out}\approx\mathcal{N}_\text{seed}$. For a BEC with $N=3\times 10^4$, 2\% seeding corresponds to a measurement of 600 atoms, which is experimentally less susceptible to noise than an average near zero. Another advantage is that the time required to perform an experiment is significantly decreased, since the dynamics proceed more quickly. The unseeded case takes $t_\text{SMD}=96$ ms to reach $\rho=0.50$, whereas the $0.2$\% seeded case takes just $t_\text{SMD}=35$ ms to reach the same fraction.

We extend these simulations to the case of double-sided seeding, which is sensitive to the initial spinor phase. In Fig. \ref{fig:AmpVsPhase}(b) we simulate the SU(1,1) interferometer with 0.1\% coherent seeds in both \mplmi\ states and plot $f_<(\phi)$ for several representative initial spinor phases $\theta$ with $t_\text{SMD}=20$ ms. Because the initial spinor phase $\theta$ is well-defined for double-sided seeding, the initial stage of the interferometer becomes a phase-sensitive amplifier, and the performance of the interferometer changes significantly with $\theta$. There are two special values of the spinor phase that for a given seed fraction separate the mean-field spinor dynamics into regions of oscillating phase and running phase \cite{Zhang:2005fk}. The phases which define this separatrix can be calculated for \ctwon$>0$ and $-2c_2\bar{n}<q<0$ using
\begin{equation}
\theta_\text{sep}=\pm\cos^{-1}\left[\frac{(\rho_++\rho_-)(1+\frac{q}{c_2\bar{n}(1-\rho_+-\rho_-)})}{\sqrt{4\rho_+\rho_-}}\right].
\end{equation}
For the 0.1\% seeds in Fig. \ref{fig:AmpVsPhase}(b) these values are $\theta_\text{sep}=\pm2.86$ rad.

One of the consequences of double-sided phase-sensitive seeding is the loss of symmetry about $\phi=0$ in the Fisher information in Fig. \ref{fig:AmpVsPhase}(b). For an average initial spinor phase $\theta=\theta_\text{sep}=-2.86$ rad, the Fisher information peaks at $f_<(\phi=-0.02)=46$, which is significantly larger than $f(0)=28$ for the unseeded case. We conclude that for a fixed spin-mixing time, the phase-sensitive input amplifier can yield a spinor interferometer with greater sensitivity than the unseeded case, depending on initial spinor phase $\theta$.

The inset to Fig. \ref{fig:AmpVsPhase}(b) plots $f_{<,\text{max}}$ versus $\theta$ with the symbols indicating the phases plotted in the main figure, and two dotted vertical lines indicating the phases of the separatrix. There are two broad peaks centered on $\theta=-2.8, 2.0$ rad that extend significantly above the unseeded scaled Fisher information $f_{<,max}=28$. In between those peaks is a broad region with decreased sensitivity and a minimum at $\theta=-0.25$ rad. The phases of the separatrix produce initial conditions corresponding to near maximum ($\theta=-2.86$ rad) and near minimum ($\theta=+2.86$ rad) Fisher information.

The effect of the spinor phase can also be illustrated for double-sided 0.1\% seeding by plotting $f_{<,\text{max}}$ versus $\rho$ (see Fig. \ref{fig:FIVsRho}(b)). We follow the performance of the interferometer by increasing $t_\text{SMD}$, which again increases $\rho$. As in the single-side seeded interferometer, we see that seeding decreases $f_{<,\text{max}}$ for a given value of $\rho$, but that for sufficient $t_\text{SMD}$ such that $\rho\rightarrow 1$, we have a scaled Fisher information much larger than the standard quantum limit. Again we find that the optimal value of $\theta$ is near to the one that defines the separatrix.

The peak scaled Fisher information for 0.1\% double-sided seeding is $f_{<,\text{max}}\approx1200$ and occurs for $\rho\approx0.47$. This value of $f_{<,\text{max}}$ is significantly larger than at the standard quantum limit and, as for the case of single-sided seeding, occurs at much earlier times than in the unseeded case. For double-sided seeding and sufficiently high gain, the sensitivity is better than that attainable using coherent state inputs in a Mach-Zehnder interferometer (dashed line in Fig. \ref{fig:FIVsRho}(b)).

\section{Experiment}

To experimentally demonstrate control of seed number and phase, we begin with a greater-than-$90\%$-pure condensate of approximately $1.4\times 10^5$ $^{23}$Na atoms in the $\ket{1,0}$ hyperfine ground state. The atoms are held in a crossed optical dipole trap with final trap frequencies of $\omega_{(x,y,z)}=2\pi(137.7(8),140.0(8),210(6))$ Hz. The indicated measurement uncertainties in this paper are one standard deviation of the mean statistical uncertainties, unless otherwise noted. The initial atomic state is purified by using microwave adiabatic rapid passage to transfer the atoms in \mplmi\ to the $F=2$ manifold, and subsequent removal from the trap with a $100\;\mu$s resonant optical pulse. In a similar way, the total atom number is reduced to $3.5\times10^4$ atoms to ensure single-mode dynamics.

\subsection{Single spatial-mode approximation}
An estimate for whether the BEC stays within the single mode approximation while undergoing spinor oscillations can be made by considering a BEC of $N$ atoms with a Thomas-Fermi radius $R_\text{TF}=\bar{a}_\text{ho}(15Na_c/\bar{a}_\text{ho})^{1/5}$, where $\bar{a}_\text{ho}=\sqrt{\hbar/(M_\text{Na}\bar\omega)}$, $M_\text{Na}$ is the atomic mass, $\bar{\omega}\approx 2\pi\times160$~Hz is the geometric mean angular trap frequency, and $a_c=(a_0+2a_2)/3=2.79(2)$~nm \cite{Knoop:2011gu}. Spin-waves can be ignored when the half-wavelength of the lowest-energy combined spatial and spin-wave is larger than $2R_\text{TF}$, or $\lambda_s/2 > 2R_\text{TF}$. The wavelength $\lambda_s=2\pi\xi_s$, where $\xi_s=1/k_s=\hbar/\sqrt{2M_\text{Na}|c_2|\bar{n}}$ is the spin-healing length of a spin-wave with wavevector $k_s$ \cite{StamperKurn:2001bu}. Within the Thomas-Fermi approximation, the single-mode condition reduces to $N_\text{SMA}<(\pi^2a_\text{ho}/(8(a_2-a_0)))^{5/4}(15a_c/a_\text{ho})^{1/4}$. The scattering length difference for sodium is $(a_2-a_0)\approx0.29$ nm \cite{Pechkis:2013vg}, giving an estimate of $N_\text{SMA}\lesssim2.6\times 10^4$ atoms for the maximum number of atoms for the validity of the SMA. Conversely, using $N=3\times 10^4$, and $c_2\bar{n}/h\approx 23$ Hz, we find a geometric mean Thomas-Fermi diameter of $2R_\text{TF}=12.6~\mu$m, and $\lambda_s/2\approx9.5~\mu$m, close to the Thomas-Fermi single-mode criterion. In order to achieve single-mode dynamics in the experiment, we carefully minimize magnetic field gradients. After this has been done, we find that additional spatial modes are not populated even for large negative $q$, where the BEC should be unstable and populate higher-order spatial modes \cite{Bucker:2012dm, Scherer:2010hs,Klempt:2010bh,Klempt:2009gw}.

\subsection{Initial phase control}
Unlike in optical four-wave mixing experiments, the single-mode atomic BEC does not have spatially distinguishable states, and we work instead with spin superposition states, where the fractions in \mpl\ and \mmi\ serve as the probe and conjugate beams respectively. First we prepare a probe state by transferring a small fraction of the \mz\ state into \mmi\ using two sequential microwave pulses: $\ket{1,0} \rightarrow \ket{2,-1}$, and then $\ket{2,-1} \rightarrow \ket{1,-1}$. We can vary the probe population between zero and 1\%. Next, the conjugate fraction is established for $m=+1$ using similar microwave pulses. For our experiments we fix the fraction in $m=+1$ at 1\%.

The initial spinor phase $\theta$ is controlled by slightly detuning one of the population transfer microwave pulses from resonance or by using a far-detuned phase-shifting microwave pulse. Figure \ref{fig:phasescan} shows our ability to set the initial phase $\theta$ with a 500 $\mu$s long microwave pulse by verifying the mean-field prediction of the spinor evolution. With single-sided seeding the final atom fraction is independent of initial phase, demonstrating phase-insensitive amplification. For seeding both \mplmi\ the amplification is phase-sensitive.

\PhaseScan

\subsection{Procedure}
The spinor amplifier is switched on with a continuous-wave off-resonant microwave field that shifts the relative energy of the levels due to the AC stark effect \cite{Gerbier:2006jt,Leslie:2009kp,Deuretzbacher:2010it,Pechkis:2013vg}. In particular, for a bias magnetic field of $B=50\ \mu$T (equivalent to a 350 kHz linear Zeeman shift), we blue-detune 150 kHz from the clock transition $\ket{1,0}\rightarrow \ket{2,0}$. The effective quadratic Zeeman shift $q$ is then
\begin{align}
\begin{split}
q&=\gamma B^2-\Delta E_0 + 0.5\left(\Delta E_{-1} + \Delta E_{+1}\right)\\
\Delta E_m&=-\sum\limits_{n}\frac{\Omega_{m,n}^2}{4\Delta_{m,m+n}}\text{ ,}
\end{split}
\end{align}
where the sum is over all microwave photon angular momentum quantum numbers $n=-1,0,1$ with Rabi frequencies $\Omega_{m,n}=\omega_n C_{m,n}$ that couple between atomic states with Clebsch-Gordan coefficients $C_{m,n}=\braket{1,m;1,n|2,m+n}$, each with detuning $\Delta_{m,m+n}$ and $\gamma/h=27.7 \text{~kHz/(mT)}^2$ \cite{Pechkis:2013vg}. The measured microwave frequencies for $\sigma^-$, $\pi$, and $\sigma^+$ coupling are respectively $\omega_{-1}= 2\pi \times 8.22(8)$~kHz, $\omega_{0}=2\pi \times 23.7(3)$~kHz, and $\omega_{+1}=2\pi \times 11.76(3)$~kHz. This mixed microwave coupling complicates the calculation of $q$, but we intentionally operate in this way so that we are able to make rapid transfers to all possible magnetic sublevels. The experimental value of $q/h$ determined by the calibrated microwave fields is $-1.3(4)$~Hz.

While the value of $q$ is stable for the duration of the experiment, \ctwon\ varies from one experimental realization to the next due to fluctuations in $N$. In order to capture these variations within the TWA simulations, we measure $N$ for each experimental realization and use it to scale the interaction energy \ctwon$=c_2\bar{n}_0(N/[3\times 10^4])^{2/5}$, where $c_2\bar{n}_0$ is the interaction energy for $3\times 10^4$ atoms. This scaling is expected from the Thomas-Fermi approximation for which $\bar{n}\propto N^{2/5}$. We also compensate for slow variations in trap frequencies by fitting each series of 35 data points to find the best value of $c_2\bar{n}_0$. The mean and standard deviation of the distribution from the fits for the entire data set are $c_2\bar{n}_0/h=22.8 \pm 0.7 \text{~Hz}$.

After the microwave dressing field is applied, the atoms are allowed to evolve for a fixed time $t_\text{SMD}=27$ ms. At that time the dipole trap is switched off, and the total number of atoms $N$ and the fractional populations $\rho_{\pm 1}$ are measured by Stern-Gerlach separation and absorption imaging after a short time-of-flight. By monitoring the fluctuations in the magnetization ($M\equiv\rho_+-\rho_-$) we estimate that our atom measurement uncertainty is 190 atoms for each spin state.

\subsection{Experimental results}
In Fig. \ref{fig:AtomAmplifier} we present experimental data on the initial phase-sensitive amplifier stage of the SU(1,1) interferometer for constant seeding into $m=+1$ of $\rho_{+,\text{seed}}=0.01$ versus the fraction seeded into $m=-1$ ($\rho_{-,\text{seed}}$). For all data, $t_\text{SMD}$ is long enough to that the system has evolved into the depleted pump regime for which the TWA simulations are necessary. The system transitions from a phase-insensitive amplifier with single-sided seeding for $\rho_{-,\text{seed}}=10^{-6}$ to a balanced double-sided phase-sensitive amplifier when $\rho_{-,\text{seed}}=0.01$. The two sets of data differ in their initial phase with $\theta\approx0$ (black circles), and $\theta=2.46$~rad (red diamonds). The initial phases become increasingly important as $\rho_{-,\text{seed}}\rightarrow\rho_{+,\text{seed}}$. The difference in the fractions at small $\rho_{-,\text{seed}}$ for the two initial phases is caused by slightly different gains produced by the somewhat larger BEC and more negative $q$ for the deamplification data.

\AtomAmplifier

For double-sided seeding with $\theta=0.016$~rad (black circles), the seeded atoms stimulate the production of more pairs of atoms, resulting in a larger amplified fraction after a fixed amount of time. For $\theta=2.45$ (red diamonds), the seeded $m=\pm1$ atoms first recombine to form pairs of $m=0$ atoms before spontaneous and stimulated emission produces new pairs of $m=\pm1$ atoms. For a fixed $t_\text{SMD}$ the time required for the initial deamplification results in a smaller measured fraction of atoms.

The transition from phase-insensitive to phase-sensitive amplifier can be understood by considering the standard deviation in the initial phase of the condensate calculated from TWA simulations (green dashed line). $\Delta\theta$ decreases as $\rho_{-,\text{seed}}$ increases, and for $\rho_{-,\text{seed}}>10^{-4}$ agrees well with the analytical expectation $\Delta\theta=\sqrt{1/\rho_++1/\rho_-+4/\rho_0}/(2\sqrt{N})$, derived from the coherent state standard deviation of the components $\Delta\theta_m=1/(2\sqrt{N_m})$. At the point that $\rho_{-,\text{seed}}=10^{-3}$, $\theta$ is sufficiently well-defined that the atom-fractions differ by three full standard deviations from the fraction for single-sided seeding, corresponding to less than 35 atoms seeded into \mmi. Due to spinor amplification, we achieve a sensitivity to initial numbers of atoms far below the 190 atoms per spin state measurement noise determined from the magnetization data (open circles).

\section{Conclusion}
We have used the truncated-Wigner approximation to simulate a spinor nonlinear-SU(1,1) interferometer seeded with coherent populations in the ``probe" and ``conjugate" states. In the case of single-sided seeding this realizes a phase-insensitive amplifier, while for double-sided seeding the input is a phase-sensitive amplifier. We quantify the performance of the interferometer by calculating the phase-sensitivity $\Delta\phi$, which we use to estimate the scaled Fisher information, $f(\phi)$. We have shown that coherent seeds suppress $f(\phi)$ around $\phi=0$, while retaining scaled Fisher information significantly beyond the standard quantum limit for the optimal phase shift. An advantage of seeding the amplifier is that the outputs have a larger mean and smaller variance, giving better signal-to-noise ratio for the experimental measurements.

Experimentally, we have demonstrated that seeding the \mplmi\ input states of a $^{23}$Na spinor BEC leads to an amplifier that is sensitive to probe and conjugate coherent states with less than 35 atoms or 0.1\% of the total number of atoms. Using microwave pulses we can control the initial phase of the spinor condensate and therefore the subsequent dynamics. Nonlinear spinor SU(1,1) interferometers constructed from these phase-sensitive amplifiers are potential systems for achieving sensitivity beyond the standard quantum limit.

\begin{acknowledgments}
JPW acknowledges support from the Research Corporation for Science Advancement. ZG acknowledges support from the NSF PFC at JQI. RB received support from the European Union's Seventh Framework Programme under Grant No.~PCIG-GA-2013-631002. ET acknowledges support by the US National Science Foundation, Grant No.~PHY-1506343.
\end{acknowledgments}

%

\end{document}